\documentclass[aps,preprint,superscriptaddress]{revtex4}
\usepackage{graphicx}

\newcommand{\bra}[1] {\left\langle #1 \right|}
\newcommand{\ket}[1] {\left| #1 \right\rangle}

\begin{document}

\title{Landau-Zener transitions in an open multilevel quantum system}

\author{S. Ashhab}
\affiliation{Qatar Environment and Energy Research Institute, Hamad Bin Khalifa University, Qatar Foundation, Doha, Qatar}

\date{\today}


\begin{abstract}
We consider the Landau-Zener problem for a multilevel quantum system that is coupled to an external environment. In particular, we consider a number of cases of three-level systems coupled to a harmonic oscillator that represents the external environment. We find that, similarly to the case of the Landau-Zener problem with a two-level system, when the quantum system and the environment are both initially in their ground states the probability that the system remains in the same quantum state is not affected by the coupling to the environment. The final occupation probabilities of the other states are well described by a common general principle: the coupling to the environment turns each Landau-Zener transition process in the closed system into a sequence of smaller transitions in the combined Hilbert space of the system and environment, and this sequence of transitions lasts a total duration that increases with increasing system-environment coupling strength. These results provide an intuitive understanding of Landau-Zener transitions in open multilevel quantum systems.
\end{abstract}


\maketitle

\section{Introduction}
\label{Sec:Introduction}

The Landau-Zener (LZ) problem is one of the basic paradigms in the physics of quantum systems under the influence of time-dependent Hamiltonians. Specifically, the LZ problem relates to the evolution of the system when two or more energy levels experience an avoided crossing as the external parameters are varied in time. The basic problem with two energy levels and a linear sweep of the external parameter turns out to be simple enough that an analytic solution for the dynamics can be obtained  \cite{Landau,Zener,Stueckelberg,Majorana}.

While the two-level problem is extremely valuable in understanding the dynamics of quantum systems at avoided crossings of energy levels, in many realistic problems there are more than two energy levels that (nearly) intersect each other in a certain region in parameter space. One example of such systems is nanomagnets, where experiments have shown the need to go beyond the two-level LZ model \cite{Wernsdorfer}. Recent experiments on Landau-Zener-St\"uckelberg interferometry in superconducting circuits have also involved multiple energy levels \cite{Berns,Sun,Shevchenko}. The field of adiabatic quantum computation (AQC) \cite{Farhi,Johnson,Lanting} is another area where recent experiments have shown that a large number of energy levels come close to each other at the most crucial point in the time evolution. Problems related to conical intersections in molecules, relevant to many photo-chemical reactions, also often involve multiple electronic states.

The multilevel LZ problem has been studied quite extensively in the literature \cite{Demkov1968,Carrol,Brundobler,Ostrovsky,Usuki,Demkov2000,Sinitsyn2002,Shytov,Vasilev,Kenmoe,Kiselev,Sinitsyn2014,
Sinitsyn2015a,Sinitsyn2015b,Patra}. Most studies have focused on finding special cases that allow analytic solutions. The approach in these studies is generally to identify special cases of the multilevel problem where the equations of motion can be reduced to those of the two-level LZ problem, and as a result the transition probabilities of the generalized models generally turn out to be given by products of the usual LZ transition probability $P_{\rm LZ}=\exp\{-2\pi\delta\}$, where the adiabaticity parameter $\delta=\Delta^2/(4v)$, $\Delta$ is the minimum gap at the center of the avoided crossing, and $v$ is the sweep rate (possibly rescaled to take into account the slopes of the energy levels).

As mentioned above, the two-level LZ problem can be solved analytically \cite{Landau,Zener,Stueckelberg,Majorana}. When one introduces a thermal environment, the problem becomes more complex and in general does not allow an analytical solution. Instead, numerous studies have tackled the problem using numerical calculations under a variety of assumptions and approximations \cite{Kayanuma,Gefen,Ao,Shimshony,Nishino,Pokrovsky2003,Sarandy,Ashhab2006,Lacour,Pokrovsky2007,Amin,Nalbach2009,
Nalbach2013,Dodin,Xu, Haikka,Nalbach2014,Ashhab2014,Javanbakht,Wild}. One notable exception is the special case where the system and environment are initially in their ground state, in which case the problem can be solved analytically \cite{Wubs}. One interesting and important result that one finds in this case is the fact that for longitudinal system-environment coupling the final occupation probabilities of the system's two quantum states are unaffected by the coupling to the environment, even though the environment could end up in a highly excited state depending on the various details of the problem. Results such as this one raise the question of whether there are similar results in the case of multilevel systems.

Here we treat a number of three-level LZ problems where the system of interest is coupled to a harmonic oscillator that models the external environment. Following the approach of Ref.~\cite{Ashhab2014}, we numerically solve the time-dependent Schr\"odinger equation of the large system comprising the three-level system that experiences the avoided crossing in addition to the harmonic oscillator. We find that an extension of the result concerning the environment independence of the system's final state is obtained in this case as well, and we find some general principles governing the dynamics in the open-system multilevel LZ problem.

The remainder of this paper is organized as follows: In Sec.~\ref{Sec:Hamiltonian} we describe the basic setup and introduce the corresponding Hamiltonian. In Sec.~\ref{Sec:NumericalCalculations} we describe our numerical calculations. In Sec.~\ref{Sec:Results} we present the results of these calculations and discuss the interpretation of the results. Section \ref{Sec:Conclusion} contains some concluding remarks.

\section{Model system and Hamiltonian}
\label{Sec:Hamiltonian}

We consider a multilevel quantum system with a linearly changing Hamiltonian. In other words, the Hamiltonian is given by
\begin{equation}
H_S = \hat{A} + \hat{B} t,
\label{Eq:HamiltonianClosedSystem}
\end{equation}
where $\hat{A}$ and $\hat{B}$ are time-independent operators, and the time variable $t$ goes from $-\infty$ at the initial time to $+\infty$ at the final time. At both extremes of the time variable, i.e.~when $|t|\rightarrow\infty$, the second term dominates and the energy eigenstates of the system are the eigenstates of $\hat{B}$ \cite{DegeneracyFootnote}. As a result, the energy eigenstates at the initial and final times are the same, except for the fact that their order in the energy level ladder changes. In the case of a two-level system (TLS), the Hamiltonian can always be expressed in the form
\begin{equation}
H_{\rm TLS} = -\frac{vt}{2} \hat{\sigma}_{z}-\frac{\Delta}{2} \hat{\sigma}_{x}.
\label{Eq:TwoLevelLZHamiltonianClosedSystem}
\end{equation}
The simple form of this Hamiltonian leads to the result that the probability of making a transition between the quantum states is determined by a single parameter, i.e.~the parameter $\delta$ defined in Sec.~\ref{Sec:Introduction} or equivalent combinations of $\Delta$ and $v$.

\begin{figure}[h]
\includegraphics[width=14.0cm]{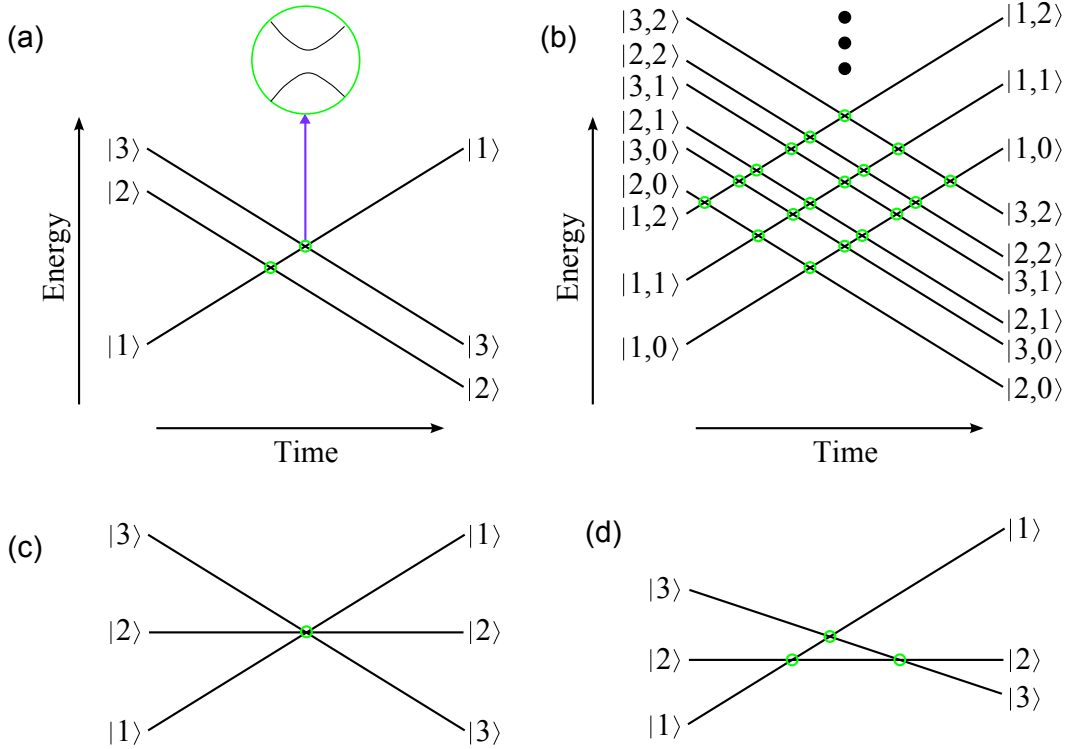}
\caption{Energy level diagrams of the three-level LZ problems considered in this paper. The three models analyzed here are the equal-slope model (a), the bow-tie model (c) and the triangle model (d). The time and energy axes, as well as the magnified avoided crossing, are not shown in Panels (c) and (d) to avoid unnecessary duplication. Panel (b) shows the energy level diagram for the equal-slope model (from Panel a) when the environment degree of freedom is included. The first number in each state label represents the state of the three-level system, while the second number represents the state of the environment starting from zero and going up to infinity. It should be noted that the environment states generally depend on the system state. For example the environment component of the state $\ket{1,0}$ is generally different from that of the state $\ket{2,0}$.}
\label{Fig:EnergyLevelDiagram}
\end{figure}

In the general case with more than two energy levels, the intermediate region can in general be a complex network of avoided crossings between all the energy levels. Even for a three-level system, the number of parameters becomes large enough that one has to consider special cases of the problem. Here we shall focus on the three-level problem and analyze the three representative cases shown in Fig.~\ref{Fig:EnergyLevelDiagram} and described in Table \ref{Table:DifferentCases}. First, we shall consider the equal-slope model, where the energy levels are divided into two groups and all the energy levels in each group have the same slope. In the present case with only three energy levels, one group will have two energy levels and the other group will have a single energy level. We shall show results only for the special case where the nonzero off-diagonal matrix elements in $\hat{A}$ are equal to each other and to the energy difference between the two parallel energy levels, but we have verified that our main results are unaffected by this specific choice, and the conclusions that we shall draw from our results should apply in the general case with any choice of parameters. Next we shall consider the bow-tie model, where the diagonal matrix elements of $\hat{A}$ are all equal to zero and all the energy levels would meet at a single point if it were not for the off-diagonal matrix elements in $\hat{A}$, which create the avoided-crossing structure. We note here that in general the bow-tie model allows quite a bit of freedom in choosing the parameters: the only requirement is that the off-diagonal matrix elements in $\hat{A}$ couple only one of the $N$ quantum states to all the other states with no direct coupling between any of these $N-1$ states. The asymptotic slopes for example do not need to satisfy any special relation. As such, we consider one special case of the bow-tie model, but we have verified that our main results are independent of this choice. Finally we shall consider the triangle model, which is in some sense the most general among three-level LZ models, because it does not have any of the symmetries contained in the other two models. The lack of any symmetry also leads to the result that there is no analytic solution for the transition probabilities in this model. As with the other two models, we consider a specific (and somewhat arbitrary) choice of parameters with the assumption that the results will contain the essential physics that is generally expected in this model.

\begin{table}[h]
\caption{The three different cases of the multilevel LZ problem analyzed in this work. Each case is defined by the operators $\hat{A}$ and $\hat{B}$, as described by Eq.~(\ref{Eq:HamiltonianClosedSystem}). As explained in the main text, the parameters used in these matrices are specific choices that we expect to capture the main features that would be obtained in general. Note that the coupling to the environment, which is described by the operator $\hat{C}$, is not included here.}
\label{Table:DifferentCases}
\begin{tabular}{cccc}
Name & $\hat{A}\times 2/\Delta$ & $\hat{B}/v$ \\ \\
Equal slope &
$\left( \begin{array}{ccc} 0 & 1 & 1 \\ 1 & 0 & 0 \\ 1 & 0 & 1 \end{array} \right)$
&
$\left( \begin{array}{ccc} 1 & 0 & 0 \\ 0 & -1 & 0 \\ 0 & 0 & -1 \end{array} \right)$
\\ \\
Bow tie &
$\left( \begin{array}{ccc} 0 & 1 & 0 \\ 1 & 0 & 1 \\ 0 & 1 & 0 \end{array} \right)$
&
$\left( \begin{array}{ccc} 1 & 0 & 0 \\ 0 & 0 & 0 \\ 0 & 0 & -1 \end{array} \right)$
\\ \\
Triangle &
$\left( \begin{array}{ccc} 0 & 1 & 0.8 \\ 1 & -2 & 0.55 \\ 0.8 & 0.55 & 0 \end{array} \right)$
&
$\left( \begin{array}{ccc} 1 & 0 & 0 \\ 0 & 0 & 0 \\ 0 & 0 & -\frac{1}{2} \end{array} \right)$
\end{tabular}
\end{table}

As mentioned above, the LZ problem with a three-level system generally contains several parameters, and defining an adiabaticity parameter is not as straightforward as it is in the case of a two-level system. However, once we specify the operators $\hat{A}$ and $\hat{B}$ with overall coefficients $\Delta$ and $v$ as shown in Table \ref{Table:DifferentCases}, we can define an overall adiabaticity parameter $\delta=\Delta^2/(4v)$. We shall use this parameter when we present our results below.

As explained in Ref.~\cite{Ashhab2014}, each method for modeling the environment in the study of the LZ problem with an open quantum system has its strengths and weaknesses. Here we use the same method as in Ref.~\cite{Ashhab2014}, i.e.~we include a harmonic oscillator that represents the external environment. With this model, we can numerically integrate the Schr\"odinger equation and obtain results whose correctness does not require making any {\it a-priori} assumptions about the dynamics. On the other hand, we might have to use logical arguments at the end to infer from our results how the system would behave under the influence of a large environment, e.g.~composed of a continuum of harmonic oscillators.

The Hamiltonian describing the multilevel LZ problem, a harmonic oscillator and coupling between the two (with the common assumption that the coupling is longitudinal and linear in the oscillator's degree of freedom) is given by:
\begin{equation}
H = \tilde{A} + \hat{B} t + \hbar \omega \hat{a}^{\dagger} \hat{a} + \hat{C} \otimes \left( \hat{a} + \hat{a}^{\dagger} \right),
\label{Eq:HamiltonianSystemPlusHarmonicOscillator}
\end{equation}
where $\omega$ is the characteristic frequency of the harmonic oscillator, and $\hat{a}$ and $\hat{a}^{\dagger}$ are, respectively, the oscillator's annihilation and creation operators. Here we assume longitudinal coupling because it leads to the intuitively natural property that away from the avoided crossings the environment only causes dephasing between the energy eigenstates of the system. It should be noted, however, that transverse coupling leads to interesting results as well, as discussed in Refs.~\cite{Wubs,Javanbakht,Demkov2000,Zueco}. The reason why we use the modified operator $\tilde{A}$ here is that the system-environment coupling causes all the energy levels corresponding to the same eigenvalue ($\ket{b_i}$) of $\hat{B}$ to be asymptotically shifted by
\begin{equation}
\Delta E_i = - \frac{\bra{b_i} \hat{C} \ket{b_i}^2}{\hbar\omega}.
\end{equation} 
The appearance of this shift can be understood by considering the last two terms in Eq.~(\ref{Eq:HamiltonianSystemPlusHarmonicOscillator}): depending on the state of the three-level system, the environment's ground state energy is shifted down by $\Delta E_i$, which in turn acts as an effective shift in the energy levels of the three-level system. In order to correct for this shift, we define the operator
\begin{equation}
\tilde{A} = \hat{A} + \sum_i \ket{b_i} \frac{\bra{b_i} \hat{C} \ket{b_i}^2}{\hbar\omega} \bra{b_i}
\end{equation}
and use it in the total Hamiltonian instead of using the original operator $\hat{A}$. If we did not make this change, the bow-tie model for example would in general turn into a triangle model because of the different shifts in the energy levels.

In order to cover several possibilities for the decoherence rates between the different quantum states, we shall use five different system-environment coupling operators in our analysis. One of these is the rather generic
\begin{equation}
\hat{C}_{1:3} = g \left( \begin{array}{ccc} 0 & 0 & 0 \\ 0 & \frac{1}{3} & 0 \\ 0 & 0 & 1 \end{array} \right),
\end{equation}
where the coefficient $g$ quantifies the overall strength of the system-environment coupling. The fact that all the diagonal matrix elements are different from each other means that the environment causes decoherence between all three quantum states of the system. The number $1/3$ is somewhat arbitrary, with the only considerations that we have taken in choosing it being that (1) we would like it to be well inside the interval (0,1) in order to cause decoherence between all three states and (2) we do not want to choose the value 1/2 in order to avoid accidental symmetries associated with the fact that 1/2 is exactly in the middle between 0 and 1. For comparison, we also perform calculations using the operator
\begin{equation}
\hat{C}_{3:1} = g \left( \begin{array}{ccc} 0 & 0 & 0 \\ 0 & 1 & 0 \\ 0 & 0 & \frac{1}{3} \end{array} \right).
\end{equation}
In the case of the equal-slope model, given that the externally tuned parameter does not affect the energetic separation between the states $\ket{2}$ and $\ket{3}$, it is quite possible that the environment might similarly not cause any fluctuations in their energetic separation and therefore not cause any significant decoherence in the subspace spanned by these two states. To treat this case, we use the operator
\begin{equation}
\hat{C}_{1:1} = g \left( \begin{array}{ccc} 0 & 0 & 0 \\ 0 & 1 & 0 \\ 0 & 0 & 1 \end{array} \right).
\end{equation}
In order to gain further insight into how the environment affects the dynamics, it is also useful to consider the two other alternatives where the environment does not decohere superpositions of two out of the three quantum states:
\begin{eqnarray}
\hat{C}_{0:1} = g \left( \begin{array}{ccc} 0 & 0 & 0 \\ 0 & 0 & 0 \\ 0 & 0 & 1 \end{array} \right),
\nonumber \\
\hat{C}_{1:0} = g \left( \begin{array}{ccc} 0 & 0 & 0 \\ 0 & 1 & 0 \\ 0 & 0 & 0 \end{array} \right).
\end{eqnarray}

Having described the different combinations of operators and parameters that we use in our analysis, next we describe further details about the numerical calculations.

\section{Numerical calculations}
\label{Sec:NumericalCalculations}

We solve the time-dependent Schr\"odinger (or Liouville-von Neumann) equation using numerical integration with the Hamiltonian given in Eq.~(\ref{Eq:HamiltonianSystemPlusHarmonicOscillator}). We use a Hilbert space constructed from product states of the system states and the lowest fifty Fock states in the harmonic oscillator. Hence we use a 150-dimensional Hilbert space in our calculations. We have verified that increasing the size of the Hilbert space to 210 does not affect the results presented here.

We have set $\hbar\omega$ to different values around $\Delta$ in different calculations, and we find that the results are generally unaffected by the exact value. All the results that presented below were obtained with $\hbar\omega=1.2\Delta$. It should be noted here that the effects of the environment on the LZ problem can be divided into two different types depending on the frequency of the environment or external noise. This distinction is generally similar to that encountered in the study of decoherence in an undriven quantum system, namely low-frequency noise causing dephasing and noise components that are resonant with specific transitions causing relaxation. More detailed discussions of these questions can be found in Refs.~\cite{Ashhab2006,Nalbach2013,Nalbach2014,Ashhab2014}. With this consideration in mind, our calculations are well suited to capture the effects of resonant frequency components but could miss some low-frequency-related effects. As discussed in Ref.~\cite{Ashhab2014} setting $\hbar\omega$ to a small value would require us to keep a large Hilbert space, which could lead to extremely long computation times. We shall therefore not perform such calculations here.

Our simulations are all started with the system and environment initialized in their ground states and the time variable set to $vt=-100\Delta$, which is sufficiently large that the system and environment barely experience any effect of the avoided crossings in the initial stages of the evolution. We evolve the time-dependent Schr\"odinger equation from this initial time to the final time given by $vt=100\Delta$. This latter value is sufficiently large that the occupation probabilities of the different quantum states will be very close to their asymptotic values.

At the final time we calculate the occupation probabilities of the three system states. Because we choose a large value of the final time, each energy eigenstate of the large system comprising the three-level system and the harmonic oscillator lies almost completely in the subspace that corresponds to one of the eigenstates of the operator $\hat{B}$, making the classification of the energy eigenstates based on the $\hat{B}$ eigenstates straightforward. It should be noted that there is no such simple correspondence at intermediate times when the energy levels are undergoing avoided crossings.

\section{Results}
\label{Sec:Results}

In this section we present the results of the numerical simulations described in Sec.~\ref{Sec:NumericalCalculations} for the three models described in Sec.~\ref{Sec:Hamiltonian} and a variety of system-environment coupling operators.

\subsection{Equal-slope model}

\begin{figure}[h]
\includegraphics[width=17cm]{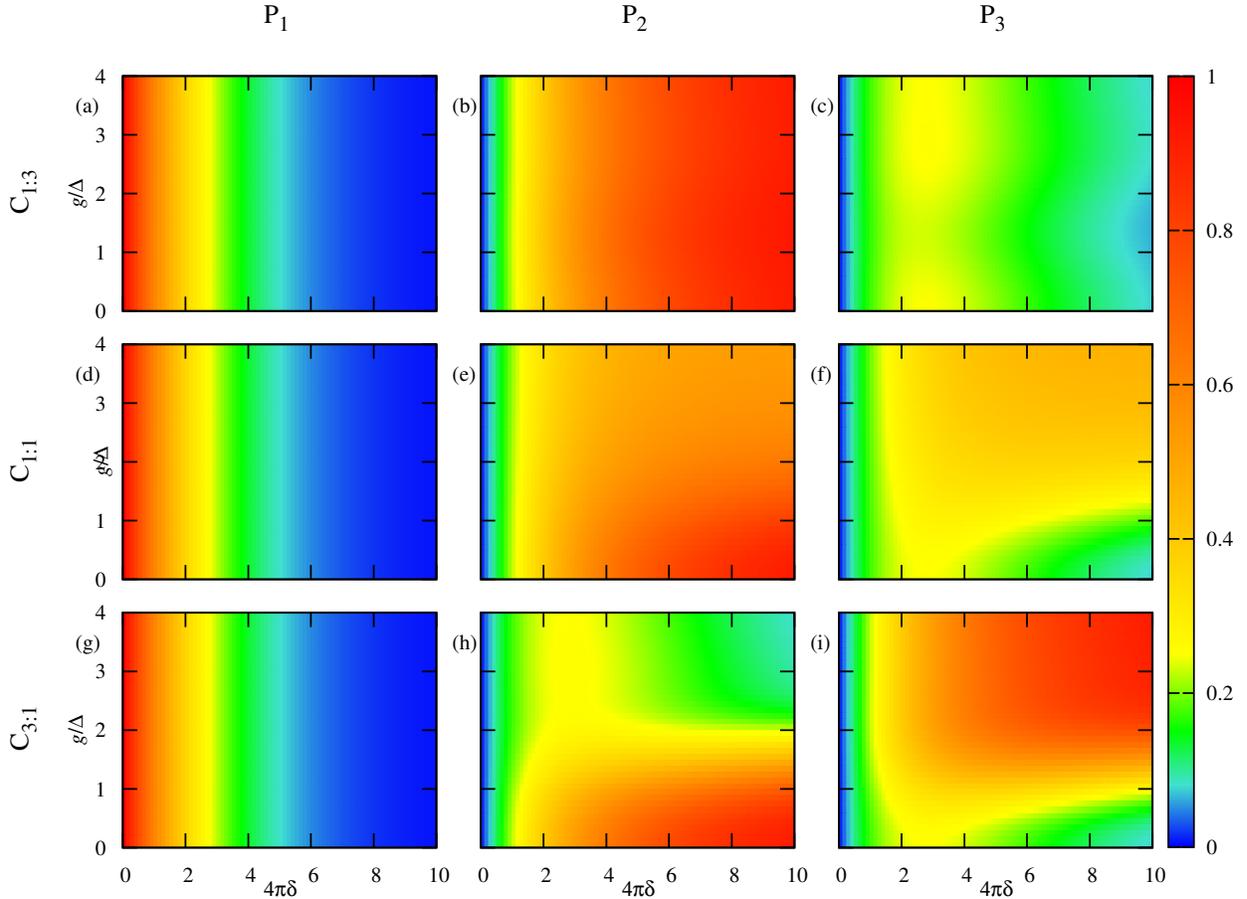}
\caption{Final occupation probabilities of the three quantum states of the system as functions of the adiabaticity parameter $\delta$ and the system-environment coupling strength $g$ for the equal-slope model. The different rows correspond to different choices of $\hat{C}$ while the different columns correspond to the three system states.}
\label{Fig:OccupationProbabilitiesEqualSlope}
\end{figure}

We start with the equal-slope model. In Fig.~\ref{Fig:OccupationProbabilitiesEqualSlope} we plot the final occupation probabilities of the different quantum states of the three-level system as functions of the adiabaticity parameter $\delta$ and the coupling strength $g$ for the three cases defined by $\hat{C}_{1:3}$, $\hat{C}_{1:1}$ and $\hat{C}_{3:1}$. In the absence of coupling to the environment, i.e.~when $g=0$, the final occupation probabilities of the states $\ket{1}$ and $\ket{2}$ generally follow the dependence seen in the two-level LZ problem: in the fast-sweep limit ($\delta\rightarrow 0$) the quantum system remains in the state $\ket{1}$, while in the adiabatic limit ($\delta\rightarrow\infty$) the system adiabatically follows the ground state and therefore ends up in the state $\ket{2}$ at the final time. As already discussed in the literature, the occupation probability of the state $\ket{3}$ vanishes in both limits, but it has a maximum value of 0.25 attained at $4\pi\delta=4\ln 2=2.77$.

When we now consider the effect of the environment, the first observation that we make from Fig.~\ref{Fig:OccupationProbabilitiesEqualSlope} is that the probability to remain in the state $\ket{1}$ is not affected by the coupling to the (zero-temperature) environment. While this result might be counter-intuitive, it is consistent with similar results that are well established in the literature, as can be seen in Refs.~\cite{Wubs,Sinitsyn2002}. We note here that to our knowledge there is no simple intuitive explanation for this result, even in the two-level case. The dependence of the occupation probabilities for the two other quantum states also contains seemingly surprising results. In the case of $\hat{C}_{1:3}$, the occupation probabilities of the states $\ket{2}$ and $\ket{3}$ are almost independent of the coupling to the environment. In fact, if we use the operator $\hat{C}_{0:1}$, which does not decohere superpositions of the states $\ket{1}$ and $\ket{2}$, the occupation probabilities become completely independent of $g$. In the case of $\hat{C}_{1:1}$, i.e.~when the environment does not decohere quantum superpositions of the states $\ket{2}$ and $\ket{3}$, the final occupation probabilities of the states $\ket{2}$ and $\ket{3}$ steadily approach each other and seem to asymptotically coincide with each other as $g$ is increased. In the case of $\hat{C}_{3:1}$ (or $\hat{C}_{1:0}$), the probability of the state $\ket{2}$ decreases and that of the state $\ket{3}$ increases with increasing $g$. Furthermore, when $g$ becomes larger than both $\Delta$ and $\hbar\omega$, the probability of the state $\ket{3}$ becomes larger than that of the state $\ket{2}$.

\begin{figure}[h]
\includegraphics[width=8cm]{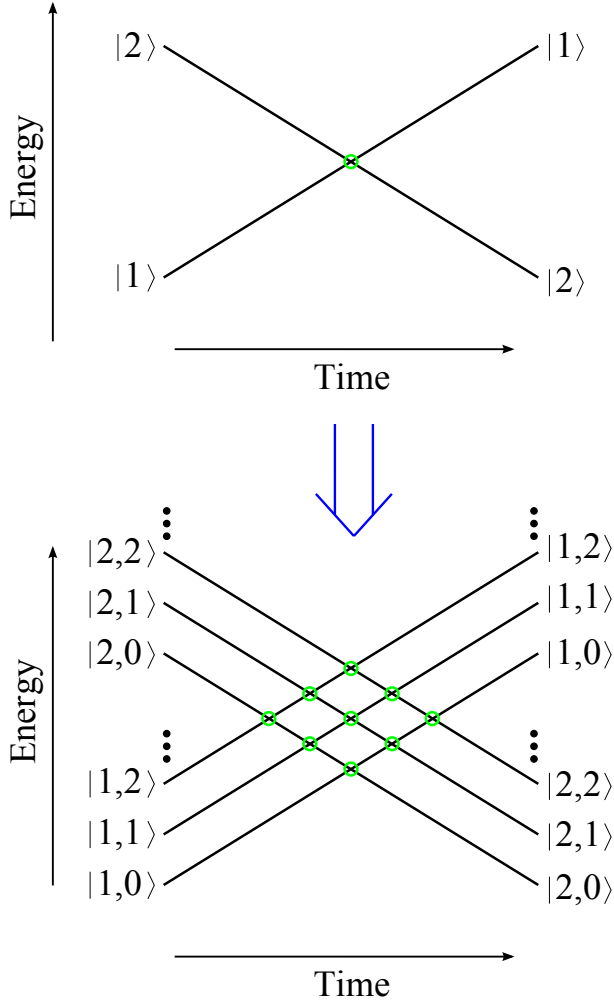}
\caption{Schematic diagram showing how the energy level structure with a single avoided crossing in the two-level LZ problem turns into a structure with an infinite number of energy levels and avoided crossings when the environment degrees of freedom are included. In particular, if the combined system is initially in the state $\ket{1,0}$, its final occupation probabilities are governed by the infinite sequence of avoided crossings between the state $\ket{1,0}$ and the states $\ket{2,0},\ket{2,1},\ket{2,2},\cdots$.}
\label{Fig:TwoLevelLandauZenerWithEnvironment}
\end{figure}

The three different types of behavior described above can be understood using a single principle that can be deduced from considering the LZ problem with a two-level system in the presence of an environment. In that case, as is illustrated in Fig.~\ref{Fig:TwoLevelLandauZenerWithEnvironment}, the single avoided crossing between the system states $\ket{1}$ and $\ket{2}$ turns into an infinite sequence of avoided crossings between the states $\ket{1,0}$ and $\ket{2,0}$, $\ket{2,1}$, $\ket{2,2}$, ..., where the second index describes the state of the environment (with the label 0 denoting the ground state of the environment, including any system-state-dependent corrections). The energy gaps of these avoided crossings follow the function $\alpha^n\exp\{-\alpha^2/2\}L_0^n(\alpha^2)/\sqrt{n!}$, where $\alpha\propto g/(\hbar\omega)$ (with the exact relation determined by the details of $\hat{C}$), $n$ is the environment state index, and $L_0^n$ are associated Laguerre polynomials, which gives a distribution peaked around $n=\alpha^2$ and with width proportional to $\alpha$ \cite{Ashhab2010}. As the system-environment coupling strength $g$ increases, the final occupation probability spreads among a larger number of final states and the population also shifts up to higher excited states in the environment. This principle can be translated to the LZ problem with a three-level system. If we consider the dynamics in the limit where the decoherence between the states $\ket{1}$ and $\ket{2}$ vanishes, some probability is transferred from the state $\ket{1,0}$ to the state $\ket{2,0}$ at their avoided crossing, and all subsequent dynamics involves transfers of probability from the state $\ket{1,0}$ to states of the form $\ket{3,n}$ with increasing values of $n$. The total probability transferred to the states $\ket{3,n}$ will be independent of $g$, as occurs in the case of a two-level system. In the opposite case where the environment does not decohere superpositions between the states $\ket{1}$ and $\ket{3}$, the single avoided crossing between these states is preserved, while an increasing value of $g$ splits the $\ket{1}$-$\ket{2}$ avoided crossing into a large number of avoided crossings most of which occur later in time than the $\ket{1}$-$\ket{3}$ avoided crossing. As a result, after a tiny transfer from the state $\ket{1}$ to the state $\ket{2,0}$, some probability is transferred from the state $\ket{1}$ to the state $\ket{3}$ according to the adiabaticity parameter of that avoided crossing traversal, and the significant part of the transfer of probability from the state $\ket{1}$ to states of the form $\ket{2,n}$ starts later in time. One consequence of this picture is that for sufficiently large values of $\delta$, and assuming a sufficiently large value of $g$, essentially all the probability will end up in the state $\ket{3}$ before any transfer from $\ket{1}$ to $\ket{2}$ has a chance to occur. In the strict adiabatic limit ($\delta\rightarrow\infty$, or more specifically $\delta\times\exp\{-(g/\hbar\omega)^2\}\gg 1$), the combined system will follow its ground state and end up in the state $\ket{2,0}$, but the exponential function in the inequality makes achieving this limit very difficult for strong system-environment coupling.  In the case of $\hat{C}_{1:1}$, and assuming a large value of $g$, the occupation probability experiences a large number of small transfers to states that alternate between being in the manifold of the state $\ket{2}$ and being in the manifold of the state $\ket{3}$. This picture explains the slow dependence on $g$ and the result that at very large values of $g$ the states $\ket{2}$ and $\ket{3}$ have equal final probabilities. Obviously, if the $\ket{1}$-$\ket{2}$ and $\ket{1}$-$\ket{3}$ coupling strengths in the matrix $\hat{A}$ were unequal, the states $\ket{2}$ and $\ket{3}$ would end up with probabilities that are proportionate with the respective coupling strengths as described by the LZ formula.

We note here that if we change the distance between the energies of the states $\ket{2}$ and $\ket{3}$, the above picture remains largely unchanged. One difference is that a larger energy difference will require a larger value of $g$ for any given feature to appear.

\subsection{Bow-tie model}

\begin{figure}[h]
\includegraphics[width=17cm]{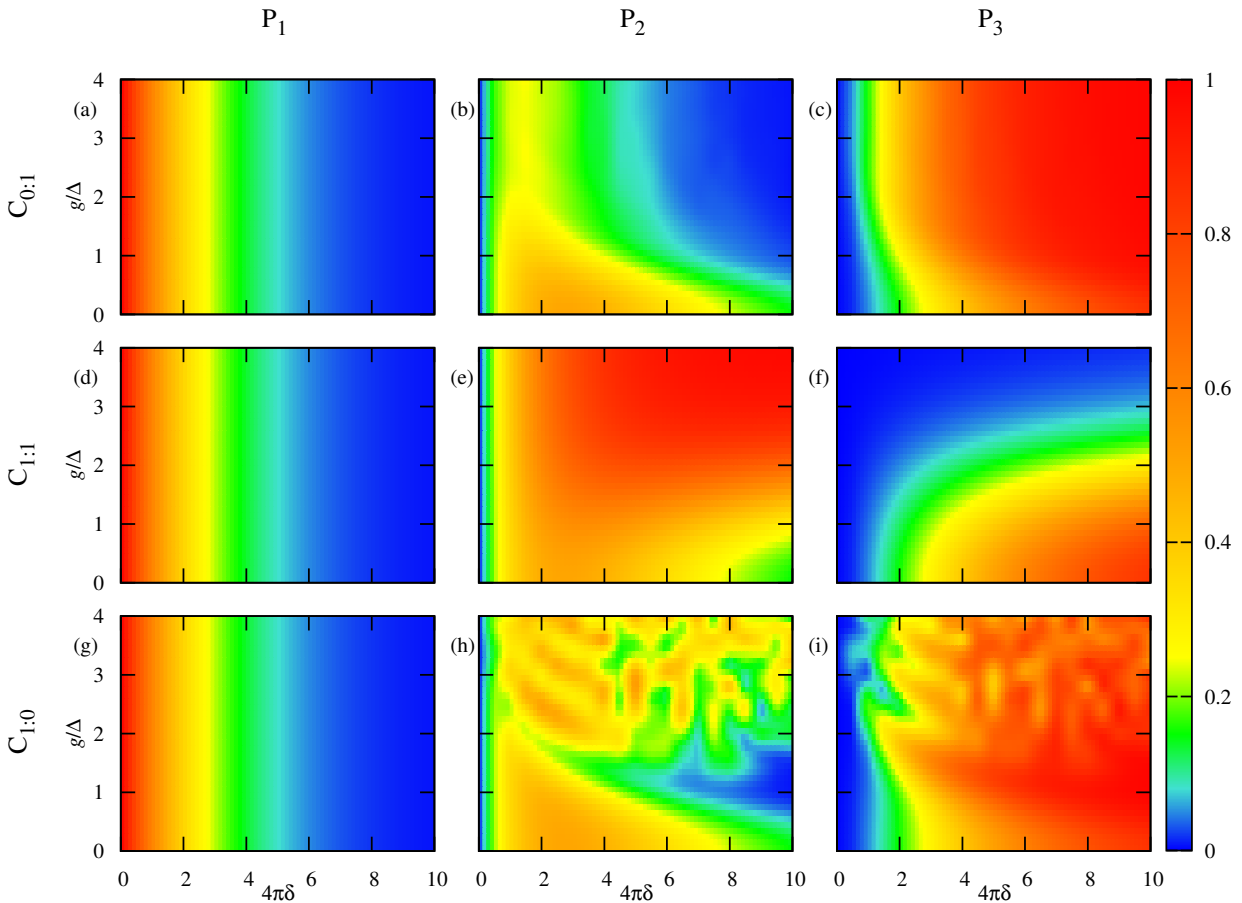}
\caption{Same as Fig.~\ref{Fig:OccupationProbabilitiesEqualSlope}, but for the bow-tie model.}
\label{Fig:OccupationProbabilitiesBowTie}
\end{figure}

\begin{figure}[h]
\includegraphics[width=8cm]{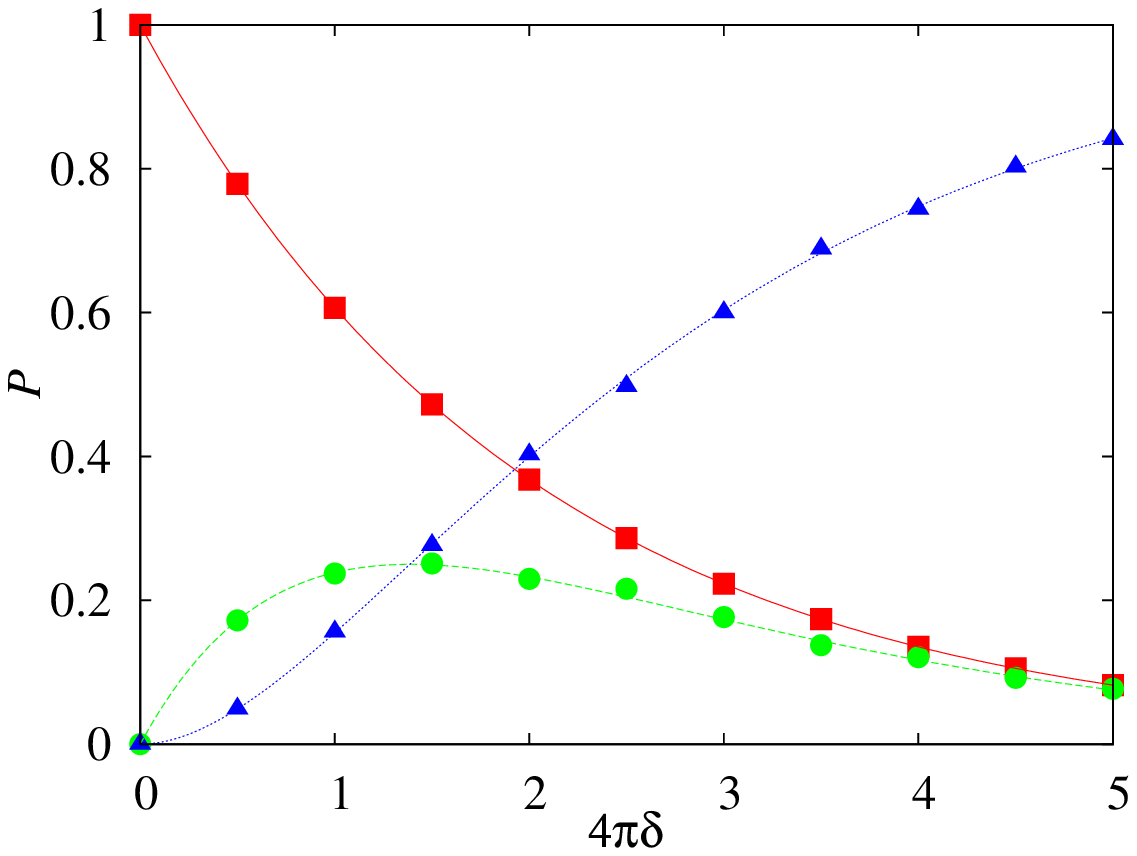}
\caption{Final occupation probabilities of the states $\ket{1}$ (red squares), $\ket{2}$ (green circles) and $\ket{3}$ (blue triangles) for the bow-tie model with coupling to the environment described by the operator $\hat{C}_{0:1}$ and strong decoherence ($g/\Delta=4$). The solid, dashed and dotted lines are the respective theoretical fits obtained by assuming that the $\ket{1}$-$\ket{2}$ avoided crossing is traversed first and is then followed by the $\ket{2}$-$\ket{3}$ avoided crossing. The theoretical formulae clearly give very good fits to the simulation data.}
\label{Fig:OccupationProbabilitiesBowTie01StrongDecoherence}
\end{figure}

We now turn to the bow-tie model. As mentioned above, we set all the non-zero matrix elements in $\hat{A}$ to the same value (as described in Sec.~\ref{Sec:Hamiltonian}), but our main results are independent of this specific choice. The final occupation probabilities for three choices of $\hat{C}$ are plotted in Fig.~\ref{Fig:OccupationProbabilitiesBowTie}.

In the absence of the coupling to the environment, it is now the state $\ket{2}$ whose population vanishes in both the limits $\delta\rightarrow 0$ and $\delta\rightarrow\infty$, with a maximum value of 0.5 attained when $4\pi\delta=4\ln 2=2.77$. In the fast-sweep limit, the system remains in the state $\ket{1}$, while in the adiabatic limit the system ends up in the state $\ket{3}$, which is the ground state at $t\rightarrow\infty$. If we now include coupling to the environment using the operator $\hat{C}_{0:1}$, the results are rather simple and match intuitive expectations based on the picture described in Sec.~\ref{Sec:Results}.A. If $g$ is very large, the avoided-crossing structure separates into two stages, first an avoided crossing between the states $\ket{1,0}$ and $\ket{2,0}$ (with transition probability $P_{\ket{1,0}\rightarrow\ket{2,0}}=1-e^{-2\pi\delta}$) followed by a sequence of avoided crossings between the state $\ket{2,0}$ and states of the form $\ket{3,n}$ with all integer values for $n$. As we have mentioned above, it is established in the literature that a sequence of this kind (i.e.~the sequence of avoided crossings between $\ket{2,0}$ and $\ket{3,n}$) gives final probabilities that are independent of the coupling to the environment, i.e.~$P_{\ket{2,0}\rightarrow\ket{2,0}}=e^{-2\pi\delta}$. As a result, the final occupation probabilities are given by $P_1=e^{-2\pi\delta}$, $P_2=(1-e^{-2\pi\delta})\times e^{-2\pi\delta}$ and $P_3=(1-e^{-2\pi\delta})\times (1-e^{-2\pi\delta})$. As shown in Fig.~\ref{Fig:OccupationProbabilitiesBowTie01StrongDecoherence}, the simulation results agree very well with the analytical formulae. The situation becomes more complicated for other choices of $\hat{C}$. For the choices  $\hat{C}_{1:3}$, $\hat{C}_{3:1}$ and $\hat{C}_{1:0}$, we find interference patterns as a result of the (generally many) different possible paths that the system can take to reach any given final state. These paths differ by the number of environmental excitations, involving states of the form $\ket{2,n}$ and $\ket{3,m}$ with all the possible values for $n$ and $m$. The interference generally results in non-monotonic dependence on both $\delta$ and $g$. In the case of $\hat{C}_{1:1}$, the dependence is monotonic, giving a suppression of the final occupation probability of the state $\ket{3}$ with increasing system-environment coupling strength $g$. Looking at the different panels in Fig.~\ref{Fig:OccupationProbabilitiesBowTie}, we can clearly see that in spite of this highly nontrivial dependence on the system parameters the probability of the state $\ket{1}$ remains unaffected by the coupling to the environment.

The bow-tie model involves direct coupling between only one quantum state and each one of the other quantum states, because this simplification allows one to obtain analytic expressions for the final occupation probabilities. We do not need to restrict ourselves to this constraint, and we can consider a generalized bow-tie model where there are off-diagonal matrix elements in $\hat{A}$ that directly couple all the eigenstates of $\hat{B}$ to each other. We have performed calculations for this case. The overall conclusions are similar to those presented above, and we do not show them in detail here. There are only two main differences that we mention: (1) an interference pattern is obtained even in the case of $\hat{C}_{0:1}$, because the direct coupling between the states $\ket{1}$ and $\ket{3}$ results in the possibility of multiple paths leading to the same final state even in this case, and (2) we do not have the suppression of state $\ket{3}$'s final occupation probability in the case of $\hat{C}_{1:1}$ anymore, because that phenomenon was the result of an interference specific to the exact parameters of the symmetric bow-tie model.

\subsection{Triangle model}

\begin{figure}[h]
\includegraphics[width=17cm]{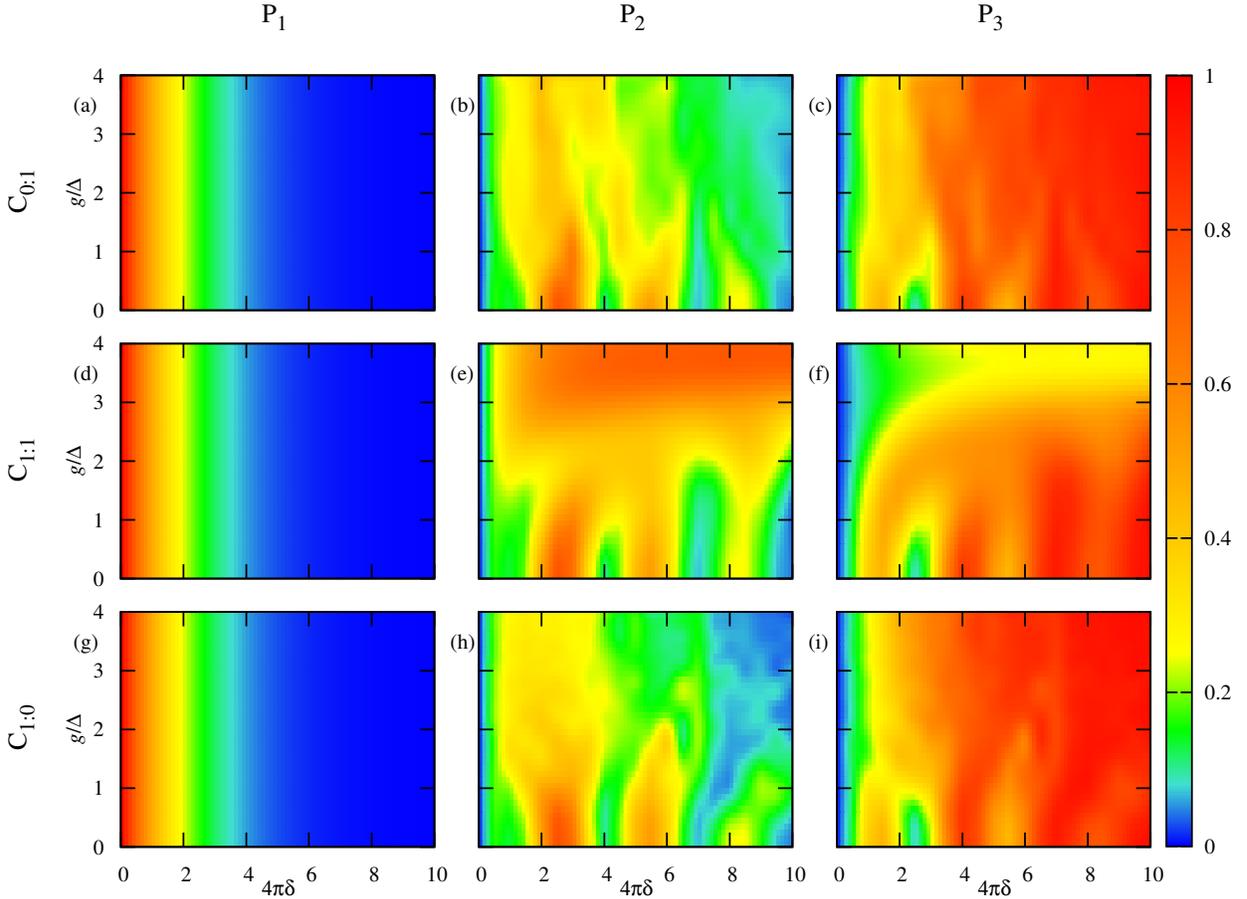}
\caption{Same as Fig.~\ref{Fig:OccupationProbabilitiesEqualSlope}, but for the triangle model.}
\label{Fig:OccupationProbabilitiesTriangle}
\end{figure}

\begin{figure}[h]
\includegraphics[width=8cm]{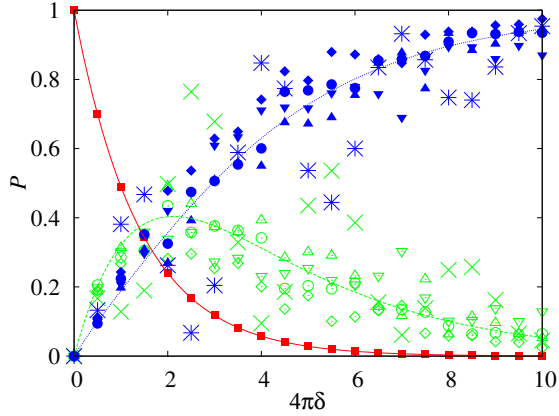}
\caption{Final occupation probabilities of the states $\ket{1}$ (red squares), $\ket{2}$ (green open symbols) and $\ket{3}$ (blue closed symbols) for the triangle model with strong decoherence ($g/\Delta=4$). More specifically the symbols are: circles for $\hat{C}_{0:1}$, triangles for $\hat{C}_{1:3}$, inverted triangles for $\hat{C}_{3:1}$ and diamonds for $\hat{C}_{1:0}$. For comparison we also show the occupation probabilities for the case $g=0$: green x symbols for the state $\ket{2}$, blue stars for the state $\ket{3}$ and none for the state $\ket{1}$ because the probabilities coincide with the those shown by the red squares. The solid, dashed and dotted lines show the respective theoretical predictions obtained by assuming that the final occupation probabilities are determined by the sequence of three LZ processes at the three avoided crossings with no contribution from quantum interference terms. The blue symbols and the green symbols both exhibit rather irregular oscillations as functions of $\delta$, and these oscillations also vary depending on the choice of $\hat{C}$. While the details of these features cannot be explained easily without a quantitative analysis, the points to note in this figure are that the theoretical formulae (dashed and dotted lines) give good overall fits to the respective sets of simulation data and in particular that decoherence suppresses the quantum-interference-induced deviations from the results of the probability-based calculation. This behavior confirms that the main effect of the coupling to the environment is to suppress the quantum-interference effects between the three LZ processes.}
\label{Fig:OccupationProbabilitiesTriangleStrongDecoherence}
\end{figure}

Next we consider the triangle model with the parameters given in Table \ref{Table:DifferentCases}. The results are plotted in Fig.~\ref{Fig:OccupationProbabilitiesTriangle}. In the absence of the coupling to the environment, a clear interference pattern is seen in the final occupation probabilities. The reason is obvious: for two of the three quantum states there are two different paths to reach the same final state, and the relative phase between the two paths depends on the different system parameters. When we include coupling to the environment, we obtain different patterns for the different choices of the operator $\hat{C}$. The general trend, however, is the same in all cases. As $g$ increases, the effects of interference between different possible paths diminishes, and the final occupation probabilities slowly approach values that are independent of $g$ or the specific choice for $\hat{C}$. In that limit, the final occupation probabilities are given by the sums of probabilities corresponding to different paths (as given by the LZ formula) without any quantum interference terms. In other words, the final occupation probabilities are given by
\begin{eqnarray}
P_1 & = & e^{-2\pi\delta}\times e^{-2\pi\delta\times 0.8^2/1.5}, \nonumber \\
P_2 & = & (1-e^{-2\pi\delta})\times e^{-2\pi\delta\times 0.55^2/0.5} + e^{-2\pi\delta}\times (1-e^{-2\pi\delta\times 0.8^2/1.5})\times (1-e^{-2\pi\delta\times 0.55^2/0.5}), \nonumber \\
P_3 & = & e^{-2\pi\delta}\times (1-e^{-2\pi\delta\times 0.8^2/1.5})\times e^{-2\pi\delta\times 0.55^2/0.5} + (1-e^{-2\pi\delta})\times (1-e^{-2\pi\delta\times 0.55^2/0.5}),
\end{eqnarray}
where the factors 0.8, 0.55, 1.5 and 0.5 are taken from Table \ref{Table:DifferentCases}. This result is illustrated in Fig.~\ref{Fig:OccupationProbabilitiesTriangleStrongDecoherence} and agrees with the intuitive picture that with strong decoherence one can think of the dynamics in terms of probability transfers between the different quantum states with no interference terms. Both Figs.~\ref{Fig:OccupationProbabilitiesTriangle} and \ref{Fig:OccupationProbabilitiesTriangleStrongDecoherence} show that, once again, the final occupation probability of the state $\ket{1}$ is independent of the coupling to the environment.

\section{Conclusion}
\label{Sec:Conclusion}

We have analyzed several instances of the LZ problem with a three-level system coupled to a harmonic oscillator that represents an uncontrolled environment. Our results all support the picture that coupling to the environment spreads out any given LZ process into a sequence of smaller LZ processes that take a longer time to be completed, albeit with a certain law that preserves the net probability transfer as long as there are no other avoided crossings being traversed at the same time. The spreading of each LZ process into multiple smaller processes means that the coupling to the environment can give rise to additional quantum interference effects in the problem, at the same time that this coupling gradually suppresses interference effects related to the parameters of the system alone. Our results also show that, assuming that the system starts in its ground state and the environment is at zero temperature, the final occupation probability of the state $\ket{1}$ is independent of the coupling to the environment. These results help enhance our understanding of the mechanisms governing multilevel LZ processes in open quantum systems and could therefore be relevant to practical applications such as adiabatic quantum computing and energy storage.

We would like to thank K.~Saito, S.~Shevchenko and M. Wubs for useful discussions.


\begin{thebibliography}{99}
\bibitem{DegeneracyFootnote} Two points should be noted here: (1) If $\hat{B}$ is not diagonal, it can be diagonalized in order to put it in simple and intuitive form at infinite times. (2) If $\hat{B}$ happens to have degenerate eigenvalues, the degeneracy in this subset of the Hilbert space can generally be lifted by defining new basis states from superpositions of the degenerate eigenstates such that the new states diagonalize $\hat{A}$ within the degenerate subspace.
\end{thebibliography}
\end{document}